\documentclass{PoS}

\title{Large-$N$ Wilsonian beta function in $SU(N)$ Yang-Mills theory by localization on the fixed points of a semigroup
contracting the functional measure}

\ShortTitle{Large-$N$ beta function and localization on fixed points}

\author{\speaker{Marco Bochicchio}\\
        INFN-Roma1 and SNS-Pisa \\
Dipartimento di Fisica, Universita' di Roma `La Sapienza' \\
Piazzale Aldo Moro 2 , 00185 Roma  \\
       E-mail: \email{marco.bochicchio@roma1.infn.it}}

\abstract{In a certain (non-commutative) version of large-$N$ $SU(N)$ $YM$ theory there are special Wilson loops,
called twistor Wilson loops for geometrical reasons, whose v.e.v. is independent on the parameter $\lambda$
that occurs in their operator definition. There is a semigroup that acts on the parameter $\lambda$ by rescaling and
on the functional measure, resolved into anti-selfdual orbits by a non-supersymmetric version of the Nicolai map,
by contracting the support of the measure.
As a consequence the twistor Wilson loops are localized on the fixed points of the semigroup of contractions.
There is a hyper-Kahler reduction on the gauge orbits of the connections with regular singularities, dense in function space in the sense of distributions 
in a neighborhood of the fixed points, that is induced by the non-supersymmetric Nicolai map.
The fixed points in the dense hyper-Kahler reduction are represented by a lattice of surface operators with holonomy in $Z_N$, i.e. by
$Z_N$ magnetic vortices. 
This localization is a non-supersymmetric analogue of the localization that occurs in the Nekrasov
partition function of the $\cal{N}$ $=2$ $SUSY$ $YM$ theory on the fixed points of a certain
torus action on the moduli space of (non-commutative) instantons.
One main consequence of the localization in the large-$N$ $YM$ case, as in the $\cal{N}$ $=2$ $SUSY$ $YM$ case, is that the beta function of the Wilsonian
coupling constant in the anti-selfdual variables is one-loop exact. Consequently the large-$N$ $YM$ canonical beta function has a $NSVZ$ form that reproduces the first two universal perturbative coefficients.}
\FullConference{The XXVIII International Symposium on Lattice Field Theory, Lattice2010\\
		 June 14-19, 2010\\
		 Villasimius, Italy}

\begin{document}

% Shorthands for \begin=={equation} and the like
 
\def\beq{\begin{equation}}
\def\eeq{\end{equation}}
\def\bea{\begin{eqnarray}}
\def\eea{\end{eqnarray}}
\def\bq{\begin{quote}}
\def\eq{\end{quote}}

\section{Introduction}

In the Duistermaat-Heckman ($DH$) localization \cite{DE} the integral of the exponential of the hamiltonian of a torus action on a compact symplectic manifold is computed exactly by the saddle-point method
as a sum on the fixed points of the torus action. The $DH$ localization has a cohomological interpretation \cite{A} according to which
the exponential of a closed form is deformed by adding a coboundary that does not change the integral of the closed form. The coboundary is rescaled by a large factor
and the modified integral is evaluated exactly by the saddle point method. Witten localization \cite{W1} extended the $DH$ idea to certain supersymmetric functional integrals that are viewed as cohomology classes for which the differential is a twisted supercharge. Nekrasov \cite{N} reproduced the Seiberg-Witten solution \cite{W3} for the prepotential  of $ \cal{N}$ $=2$ $SUSY$ $YM$ 
by localizing the partition function, i.e. the "cohomology of $1$", whose logarithm is the prepotential,
on a sum of finite dimensional integrals over the moduli space of instantons \cite{N}. 
There is a torus action, $U(1)^{N-1} \times U(1) \times U(1)$, on this moduli space compactified by a non-commutative deformation of space-time \cite{N}.
$U(1)^{N-1}$ is the unbroken gauge group of $ \cal{N}$ $=2$ $SUSY$ $YM$ in the phase spontaneously broken by the condensation of the eigenvalues of a complex Higgs field and $U(1) \times U(1)$ is the  rotational symmetry in the non-commutative background.  
As a consequence Nekrasov further reduced the partition function to a sum on the fixed points of the aforementioned torus action on the instantons moduli by means of the $DH$ localization \cite{N}. 
Nekrasov exact localization on a sum of saddle points has the interesting consequence that the beta function for the Wilsonian coupling constant
of $ \cal{N}$ $=2$ $SUSY$ $YM$ is one-loop exact, a result already known \cite{AA}. The Wilsonian coupling occurs in the Wilsonian normalization of the action as opposed to the canonical coupling that occurs by rescaling the kinetic term in such a way that is independent on the coupling. The two definitions of the coupling have different beta functions in general  \cite{AA}.
In the pure $SU(N)$ $YM$ theory there is no natural cohomology because of the lack of supersymmetry.
However, we may wonder as to whether a different kind of localization holds, perhaps linked to the large-$N$ limit.
It is clear that such localization may exist only for special observables, since this is already the case in $SUSY$ gauge theories. 
The aim of this paper is to show that a new kind of localization holds in the large-$N$ limit of pure $YM$.
In pure $YM$  in absence of a natural cohomology we developed new localization techniques based on homology theory and on a new holomorphic version of the loop equation for special Wilson loops \cite{MB}, called twistor Wilson loops for geometrical reasons that we do not discuss here,
whose v.e.v. is invariant at quantum level for deformations that are vanishing boundaries in homology (i.e. backtracking arcs). In fact twistor Wilson loops are trivial in the large-$N$ limit and in the technical sense of being invariant for the addition of backtracking arcs they are in the "homology of $1$ ".
However, the localization of the holomorphic loop equation by homology admits a simpler interpretation
directly in terms of the functional integral, as localization on the fixed points of a semigroup contracting the functional measure and leaving invariant
the v.e.v. of twistor Wilson loops. This new localization has many analogies with Nekrasov localization on the fixed points of the torus action
on the moduli of (non-commutative) instantons.
Thus it is this simpler theory that we describe in the following section.
As a consequence the large-$N$ Wilsonian beta function of $YM$, in certain new variables of anti-selfdual ($ASD$) type that are defined through a non-$SUSY$ version \cite{MB} of the Nicolai map 
\cite{Nic} of $ \cal{N}$ $=1$ $SUSY$ $YM$, that was worked out by De Alfaro-Fubini-Furlan-Veneziano in the eighties \cite{V}, is one-loop exact as in $SUSY$ gauge theories.

\section{Localization in pure large-$N$ $YM$ theory on the fixed points of a semigroup of contractions}

We define twistor Wilson loops in the $YM$ theory
with gauge group $U(N)$ on $R^2 \times R^2 _{\theta}$ with complex coordinates $(z=x_0+i x_1, \bar z=x_0-i x_1, u=x_2+i x_3, \bar u=x_2-i x_3)$ 
and non-commutative parameter $\theta$, satisfying $ [\partial_u, \partial_{\bar u}]=\theta^{-1} 1$, (see \cite{MA,AG} for a review of non-commutative gauge theories)
as follows:
\bea
Tr \Psi_{\lambda}(L_{ww})=Tr P \exp i \int_{L_{ww}}(A_z+\lambda D_u) dz+(A_{\bar z}+ \lambda^{-1}D_{\bar u}) d \bar z ,
\eea
where $D_u=\partial_u+i A_u$ is the covariant derivative along the non-commutative direction $u$. The plane $(z, \bar z)$ is instead commutative. The loop, $L_{ww}$, starts and ends at the marked point, $w$. The trace in Eq.(2.1)
is over the tensor product of the $U(N)$ Lie algebra and of the infinite dimensional Fock space that defines
the Hilbert space representation of the non-commutative plane $(u, \bar u)$ \cite{MA,AG}. The limit of infinite non-commutativity in the plane $(u, \bar u)$ is understood, being equivalent to the large-$N$ limit of the commutative gauge theory \cite{MA,AG}. Therefore non-commutativity is for us just a mean to define the large-$N$ limit as well as it is for Nekrasov just a mean to compactify the moduli space of instantons.
It easy to prove that the v.e.v. of the twistor Wilson loops is independent on the parameter $\lambda$:
\bea
<\frac{1}{N}Tr\Psi_{\lambda}(L_{ww})>=<\frac{1}{N}Tr\Psi_{1}(L_{ww})> .
\eea
The proof is obtained changing variables, rescaling functional derivatives in the usual definition of the functional integral
of the non-commutative $YM$ theory \cite{MA,AG}.  The formal non-commutative integration measure is invariant under such rescaling because of the
pairwise cancellation of the powers of $\lambda$ and $\lambda^{-1}$. The non-commutative $YM$ action, proportional to
$Tr( -i[ D_{\alpha},  D_{\beta}]-\theta^{-1}_{\alpha \beta})^2$, is invariant because of rotational invariance in the non-commutative plane. The only possibly dangerous terms couple the non-commutative
parameter to the commutator $[ D_{u},  D_{\bar u}]$, while all the other mixed terms are zero in our case. But the commutator is invariant under $\lambda$-rescaling.
In fact the twistor Wilson loops are trivially $1$ at large-$\theta$ to all orders in the 't Hooft coupling constant $g$:
\bea
\lim_{\theta \rightarrow \infty} <\frac{1}{N}Tr\Psi_{\lambda}(L_{ww})>=1 .
\eea
We do not give a diagrammatic proof of the triviality in this paper,
but we show that indeed triviality holds to the lowest non-trivial order in perturbation theory.
We have in the Feynman gauge in the large-$\theta$ limit:
\bea
<\int_{L_{ww}}(A_z+\lambda D_u) dz+(A_{\bar z}+ \lambda^{-1}D_{\bar u}) d \bar z \int_{L_{ww}}(A_z+\lambda D_u) dz+(A_{\bar z}+ \lambda^{-1}D_{\bar u}) d \bar z> \nonumber \\
= 2 \int_{L_{ww}} dz\int_{L_{ww}}d \bar z (<A_z A_{\bar z}>+i^2 <A_u A_{\bar u}>) 
=0 .
\eea
We use the $\lambda$-independence to prove that the v.e.v. of twistor Wilson loops is localized on the fixed points of the semigroup
rescaling $\lambda$.
It is convenient to choose our twistor Wilson loops in the adjoint representation and to use the fact that in the large-$N$ limit
their v.e.v. factorizes in the product of the v.e.v. of the fundamental representation and of its conjugate.
Then, for the factor in the fundamental representation, localization proceeds as follows.
We write the $YM$ partition function by means of a non-$SUSY$ analogue \cite{MB} of the Nicolai map \cite{Nic} of $\cal{N}$ $=1$ $SUSY$ $YM$ theory \cite{V},
introducing in the functional integral the appropriate resolution of identity:
\bea
1= \int \delta(F^{-}_{\alpha \beta}-\mu^{-}_{\alpha \beta}) \delta\mu^{-}_{\alpha \beta} , 
\eea
\bea
Z=\int \exp(-\frac{N 8 \pi^2 }{g^2} Q-\frac{N}{4g^2} \sum_{\alpha \neq \beta} \int Tr_f(\mu^{-2}_{\alpha \beta}) d^4x)
 \delta(F^{-}_{\alpha \beta}-\mu^{-}_{\alpha \beta}) \delta\mu^{-}_{\alpha \beta} \delta A_{\alpha} .
\eea 
$Q$ is the second Chern class (the topological charge) and $\mu^{-}_{\alpha \beta}$
is a field of $ASD$ type. The equations of $ASD$ type in the resolution of identity,
$F_{01}-F_{23}=\mu^-_{01} ,
F_{02}-F_{31}=\mu^-_{02} ,
F_{03}-F_{12}=\mu^-_{03} $,
can be rewritten in the form of a Hitchin system (taking into account the central extension that occurs in the non-commutative case), $
-i F_A+[D,\bar D] -\theta^{-1}1=\mu^0=\frac{1}{2}\mu^-_{01} ,
-i\partial_{A}  \bar D= n=\frac{1}{4}(\mu^-_{02}+i\mu^-_{03}) ,
-i\bar \partial_A D=\bar n=\frac{1}{4}(\mu^-_{02}-i\mu^-_{03}) $,
or equivalently in terms of the non-hermitian
connection whose holonomy is computed by the twistor Wilson loop with parameter $\rho$,
$B_{\rho}=A+\rho D+ \rho^{-1} \bar D=(A_z+ \rho D_u) dz+(A_{\bar z}+ \rho^{-1} D_{\bar u}) d \bar z$:
$-i F_{B_{\rho}} -\theta^{-1}1= \mu_{\rho}=\mu^0+\rho^{-1} n- \rho \bar n ,
-i\partial_{A}  \bar D= n ,
-i\bar \partial_A D=\bar n$.
The resolution of identity in the functional integral
then reads:
\bea
1=\int \delta n \delta  \bar n \int_{C_{\rho}} \delta \mu_{\rho} \delta(-i F_{B_{\rho}} - \mu_{\rho}-\theta^{-1}1) \delta(-i\partial_{A} \bar D- n) \delta(-i\bar \partial_A D- \bar n) ,
\eea
where the measure, $\delta \mu_{\rho}$, along the path, $C_{\rho}$, is over the non-hermitian path with fixed $n$ and $\bar n$ and varying $\mu^0$. The resolution
of identity is independent, as $\rho$ varies, on the complex path of integration $C_{\rho}$. 
Let us consider the v.e.v. of twistor Wilson loops:
\bea
\int \delta n \delta \bar n \int_{C_{\rho}}\delta  \mu_{\rho} \exp(-\frac{N 8 \pi^2 }{g^2} Q-\frac{N 4}{g^2}  \int Tr_f( \mu^{0})^2  +4Tr_f(n \bar n) d^4x) \nonumber\\
 Tr_f P \exp i \int_{L_{ww}}(A_z+\lambda D_u) dz+(A_{\bar z}+ \lambda^{-1}D_{\bar u}) d \bar z  \nonumber \\
\delta(-i F_{B_{\rho}} - \mu_{\rho}-\theta^{-1}1) \delta(-i\partial_{A} \bar D- n) \delta(-i\bar \partial_A D- \bar n) 
\delta A \delta \bar A \delta D \delta \bar D
\eea
and let us change variables in the functional integral rescaling the non-commutative covariant derivatives:
\bea
\int \delta n \delta \bar n \int_{C_{\rho}}\delta  \mu_{\rho} \exp(-\frac{N 8 \pi^2 }{g^2} Q-\frac{N4}{g^2}  \int Tr_f(\mu^{0})^2+4Tr_f(n \bar n) d^4x)\nonumber \\
 Tr_f P \exp i \int_{L_{ww}}(A_z+D'_u) dz+(A_{\bar z}+ D'_{\bar u}) d \bar z  \nonumber \\
\delta( -i  F_A+ [D',\bar D']-\theta^{-1}1-\mu^0
-i\frac{\lambda}{\rho} \partial_{A}  \bar D' +i \frac{\rho}{\lambda} \bar \partial_A D' -\rho^{-1} n+\rho \bar n)\nonumber \\
\delta(-i \lambda \partial_{A} \bar D'-n) \delta(-i \lambda ^{-1}\bar \partial_A D'-  \bar n)
\delta A \delta \bar A \delta D' \delta \bar D' .
\eea
Taking the limit $\lambda \rightarrow 0$ inside the functional integral, the last line implies localization on $ n=0$ and 
$\bar \partial_A D' =0$. The $\delta n$ integral is performed by means of the delta function.
The independence on the path $C_{\rho}$ in the neighborhood of $\rho=0$, that we denote, choosing $\rho=\lambda$, $C_{0^+}$, implies that the $\delta \bar n$ integral decouples and that 
$\partial_{A}  \bar D'=0$ as well.
Indeed on $C_{0^+}$ the argument of the remaining delta function contains the combination of a hermitian $-i  F_A+ [D',\bar D']-\theta^{-1}1-\mu^0$
and an anti-hermitian $-i \partial_{A}  \bar D' +i  \bar \partial_A D'$ part, whose sum can be zero only if the two terms are zero separately, using implicitly the constraint
from the conjugate representation. Therefore $-i \partial_{A}  \bar D' +i  \bar \partial_A D'=0$ on
$C_{0^+}$ and because $\bar \partial_A D' =0$ also $\partial_{A}  \bar D'=0$.
We notice that the localized density has a holomorphic ambiguity,
since we can represent the same measure using a different density making holomorphic transformations without spoiling the localization: $
\delta  \mu_{0^{+}}= \frac{\delta  \mu_{0^{+}}} {\delta  \mu'_{0^{+}}} \delta \mu'_{0^{+}} $.
This holomorphic ambiguity (and the associated holomorphic anomaly) can be resolved only through the more refined theory
of the homological localization of the loop equation \cite{MB1} that will not be discussed here. This theory requires that $\mu'$ be chosen in the holomorphic gauge, $B_{\bar z}=0$. 
The final result for the localized effective measure is:
\bea
\big[ \int_{C_{0^{+}}}  \delta \mu'_{0^{+}}  \frac{\delta  \mu_{0^{+}} } { \delta  \mu'_{0^{+}} }
\exp(-\frac{N 8 \pi^2 }{g^2} Q-\frac{N}{4g^2} \sum_{\alpha \neq \beta} \int Tr_f(\mu^{-2}_{\alpha \beta}) d^4x)
\delta(F^{-}_{\alpha \beta}-\mu^{-}_{\alpha \beta})  \big]_{n=\bar n=0} \delta A_{\alpha} ,
\eea
where we have reintroduced the covariant notation.
A delicate point arises about the meaning of the residual complex functional integral at the fixed points.
We would like to reduce it to something more manageable. In the $SUSY$ case the original functional integral is reduced to a sum over finite dimensional integrals over instantons
by cohomological localization. Of course this is not possible in the pure $YM$ case. But the basic idea in the $YM$ theory is to reduce the resolution of identity due to the non-$SUSY$ Nicolai map in Eq.(2.5) to finite dimensional integrals
on a dense set in function space in a neighborhood of the fixed points by interpreting it as hyper-Kahler reduction on a lattice of surface operators:
\bea
1= \prod_p \int d n_p   d\bar n_p \int_{C} d\mu_p \delta(-i F_{B} -\sum_{p} \mu_p \delta^{(2)}(z-z_{p_{(u, \bar u)}}) -\theta^{-1}1)  \nonumber \\
\delta(-i\partial_{A} \bar D-\sum_{p}  n_p \delta^{(2)} (z-z_{p_{(u, \bar u)}}) )
 \delta(-i\bar \partial_A D- \sum_{p}  \bar n_p \delta^{(2)} (z-z_{p_{(u, \bar u)}})) 
\eea
Codimension-two singularities of this kind were introduced many years ago in \cite{MB2,MB3} in the pure $YM$ theory as an "elliptic fibration of parabolic bundles" for the purpose of getting control over the large-$N$ limit of the pure $YM$ theory exploiting the integrability of the Hitchin fibration.
Later, in \cite{W2}, they were introduced in the $\cal{N}$ $=4$ $SUSY$ $YM$ theory for the study of the geometric Langlands correspondence, under the name of "surface operators", and this is now the name universally used in the physics literature. In fact they have been studied in the mathematics literature at classical level in \cite{KM} as singular instantons. 
The localization on the fixed points in Eq.(2.10) implies $n_p=\bar n_p=0$.
We want to understand the residual $d\mu_p$ integration. The hyper-Kahler reduction on surface operators on non-commutative space-time is not immediately understood
because of the non-commutativity. But on a non-commutative torus, for (large) rational values of $2 \pi \theta$ in units of the torus area, we can use the Morita equivalence (see \cite{AG} for a review) of the non-commutative $U(N)$ $YM$
theory to a commutative one with larger $U(N')$ gauge group and some units of a background 't Hooft flux. The large-$N$ limit must now be taken via rational sequences because of Morita duality, but this is not restrictive for twistor Wilson loops because of their triviality. 
If the $(z, \bar z )$ plane is compactified to a sphere the standard hyper-Kahler reduction has a topological interpretation as defining a representation of the fundamental group of the punctured sphere, $\prod_p M_p = 1$, via the holonomy representation of the connection $B$, where $M_p=  P \exp i \int_{L_p} B_z dz$ is the holonomy in the holomorphic gauge, $B_{\bar z}=0$,  along vanishing cycles, $L_p$, encircling the punctures, $p$.  In the holomorphic gauge $B$ is a holomorphic connection with regular singularities.
The complexification of the global gauge group acts on the holonomy at one point, $p_1$, by the adjoint action in such a way that $M_{p_1}$can be put in canonical form.
$M_{p_1}$ can be diagonalized if it has distinct eigenvalues, while in general it can be put in Jordan form.
In the large-$N$ limit it is possible to restrict the integration measure $d\mu_p$ to orbits whose holonomies have fixed eigenvalues since this restriction
implies an error of subleading order in $\frac{1}{N}$.
In addition by translational invariance the conjugacy class of the orbits at all the points $p$ must be the same. 
Finally the global $SU(N)$ gauge group of the $YM$ theory (and not only the $U(1)^{N-1}$ torus as in the Nekrasov case) must fix $M_{p_1}$, i.e. $g M_{p_1} g^{-1}=M_{p_1}$, since otherwise $M_{p_1}$ would break spontaneously the global gauge symmetry. Therefore $M_{p_1}$
must be central and thus must be in $Z_N$. But then all the orbits collapse to a point and there are no moduli at the fixed points.
However, in any neighborhood of the fixed points of the global gauge group, defined deforming infinitesimally the eigenvalues, the orbits are non-trivial and moduli there exist.
Thus if we first compute the effective measure in a neighborhood of the fixed points and then we sit on the fixed points
the induced measure will contain the powers of the Pauli-Villars regulator due to the moduli.
This has an analogue in the localization of the $\cal{N}$ $=2$ $SUSY$ $YM$ partition function, where
generically instantons have moduli (this is essential to get the correct beta function in that case too), but the instantons at the fixed points of
the torus action have not. Thus at the fixed points the contour integral over $\mu_{0^{+}}$ collapses to a discrete sum over sectors with $Z_N$ holonomy.
Now the integration on the gauge connection in Eq.(2.10) can be explicitly performed in the Feynman gauge to obtain: 
\bea
\sum_{Z_N}  [\exp(-\frac{N 8 \pi^2 }{g_W^2} Q-\frac{N}{4g_W^2} \sum_{\alpha \neq \beta} \int Tr_f(\mu^{-2}_{\alpha \beta}) d^4x)\nonumber \\
Det^{-\frac{1}{2}}(-\Delta_A \delta_{\alpha \beta} -i ad_{ \mu^-_{\alpha \beta}} ) Det(-\Delta_A)
(\frac{\Lambda}{2 \pi})^{n_b} Det^{\frac{1}{2}} \omega    \frac{\delta  \mu_{0^{+}} } { \delta  \mu'_{0^{+}} }     \times c.c. \big] _{n=\bar n=0}.
\eea
The complex conjugate factor arises by the conjugate representation. According to Eq.(2.3) the holonomy of twistor Wilson loops at the fixed points is trivial, because of the cancellation of $Z_N$ factors 
between the fundamental and the conjugate representations. 
The connection, $A$, denotes the solution of the equation
$[F^-_{\alpha \beta}- \sum_{p} \mu ^-_{\alpha \beta}(p) \delta^2(z-z_{p_{(u, \bar u)}}) =0]_{n=\bar n=0} $ in each $Z_N$ sector.
$Det^{\frac{1}{2}} \omega$ is the contribution of the $n_b$ zero modes due to the moduli and $\Lambda$ the corresponding Pauli-Villars regulator.
Important technical issues are the control over the ultraviolet quadratic divergence of the classical action on surface operators by means of the partial non-commutative Eguchi-Kawai reduction \cite{MB, MA}
and the point splitting regularization of the loop expansion of the functional determinants in the background of the lattice of surface operators \cite{MB1}.
The beta function for the Wilsonian coupling of the large-$N$ $YM$ theory in the $ASD$ variables is exactly one-loop
and coincides with the result of one-loop perturbation theory.
However, the one-loop result for $\beta_0= \frac{1}{(4\pi)^2} \frac{11}{3}$ is obtained as the sum $\frac{1}{(4\pi)^2} (\frac{5}{3}+2)$.
The first term is the total contribution of the determinant due to the non-$SUSY$ Nicolai map and of the Faddeev-Popov determinant, and gives rise to the multiplicative renormalization factor, $Z^{-1}=1-\frac{1}{(4\pi)^2} (\frac{10}{3}) g_W^2 \log(\frac{\Lambda}{\mu})$, in the $ASD$ variables. This contribution to the beta function occurs generically around any translational invariant background of surface operators.
The second term is due to the zero modes and depends on the relative normalization between the classical action in the reduced Eguchi-Kawai theory, $\frac{2(2 \pi)^2}{2 g_W^2} k( N-k) +c.c.$, and the real dimension of the moduli space, $2k(N-k)$, of the adjoint orbit, $g_p \lambda g_p^{-1}$, in an infinitesimal neighborhood of a surface operator with $Z_N$ holonomy and of its complex conjugate.
At the same time, in the regularization scheme of the holomorphic loop equation \cite{MB},
by rescaling the fields in canonical form the following relation is obtained between the canonical and the Wilsonian coupling constant:
\bea
\frac{1}{2 g^2_W}
 = \frac{1}{2 g^2_c}+ \frac{4}{(4 \pi)^2} \log g_c + \frac{1}{(4 \pi)^2} \log Z .
\eea
Differentiating this relation it follows the canonical beta function \cite{MB} that has a structure of $NSVZ$ type and reproduces the first two universal perturbative coefficients:
\bea
\frac{\partial g_c}{\partial \log \Lambda}=\frac{-\beta_0 g_c ^3 +
\frac{1}{(4 \pi)^2} g_c ^3 \frac{\partial \log Z}{\partial  \log \Lambda} }{1- \frac{4}{(4 \pi)^2} g_c^2 }  .
\eea
\acknowledgments{We thank the organizers of the Lattice 2010 conference, and in particular Guido Martinelli and Giancarlo Rossi,
for the fruitful atmosphere, for the interesting discussions and for inviting us to give this talk.}

\end{document}